\documentclass[aps,prl,twocolumn,superscriptaddress]{revtex4}
\usepackage{epsfig,amssymb,amsmath,amsfonts}
\usepackage{color,dsfont}
\usepackage{braket}
\usepackage[normalem]{ulem} 

\def\comment#1{}
\def\labell#1{\label{#1}}
\def\togli#1{}
\begin{document}
\newcommand\unit{\mathinner{\hbox{1}\mkern-4mu\hbox{l}}}
\renewcommand{\cot}{\, \hbox{\rm cotan}}
\def\>{\rangle}\def\<{\langle}
\title{State-Independent Uncertainty Relations}
\author{Hubert de Guise}
\affiliation{Department of Physics, Lakehead University, Thunder Bay, Ontario P7B 5E1, Canada}
\author{Lorenzo Maccone}
\affiliation{Dip.~Fisica and INFN Sez.\ Pavia, University of Pavia, via Bassi 6, I-27100 Pavia, Italy}
\author{Barry C.\ Sanders}
\email{sandersb@ucalgary.ca}
\homepage{iqst.ca/people/peoplepage.php?id=4}
\affiliation{Institute for Quantum Science and Technology, University of Calgary, Calgary, Alberta, T2N 1N4, Canada}
\affiliation{Program in Quantum Information Science, Canadian Institute for Advanced Research, Toronto, Ontario M5G 1M1, Canada}
\author{Namrata Shukla}
\affiliation{Institute for Quantum Science and Technology, University of Calgary, Calgary, Alberta, T2N 1N4, Canada}
 \begin{abstract}
  The standard state-dependent Heisenberg-Robertson uncertainly-relation lower bound
   fails to capture the quintessential incompatibility of observables
   as the 
   bound can be zero for some states.  To remedy this problem,
   we establish a class of tight (i.e., inequalities are saturated)
   variance-based sum-uncertainty relations derived from the Lie algebraic properties of  
   observables and show that our lower bounds depend only on the
   irreducible representation assumed carried by the Hilbert space of
   state of the system.  We illustrate our result for the cases of the
   Weyl-Heisenberg algebra, special unitary algebras up to rank~4, and
   any semisimple compact algebra. We also prove the usefulness
     of our results by extending a known variance-based entanglement
     detection criterion.
\end{abstract}
\maketitle

For $\Delta w^2$ signifying the variance of measurement outcomes for
the observable~$w$, Heisenberg's uncertainty relation for
position $x$ and momentum $p$ is  
\begin{equation}
\label{eq:xp}
	\Delta x^2\Delta p^2\geq 1/4,
\end{equation}
where $[x,p]
		=\text{i}\mathds{1},$
	and~$\mathds{1}$ is the identity operator. Eq.~(\ref{eq:xp}) fortuitously has a constant lower bound due to the appealing
algebraic properties of the commutator of ~$x$ and~$p$.  Robertson's generalization to
$\Delta A^2\Delta B^2\geq|\left\langle[A,B]\right\rangle|^2/4$ for
arbitrary observables~$A$ and~$B$ more generally has a state-dependent lower
bound~\cite{Rob29}, and so fails to 
capture the \emph{intrinsic} incompatibility of non-commuting
observables~\cite{CCB17,SRW16}.  This 
cannot be amended as the underlying product of uncertainties is null
whenever one of the uncertainties is null, an observation that
provided impetus for the emergence of uncertainty
relations~\cite{BM75,Deu83,MU88,WW10,CP14,CBTW17} that eschew variance
in favor of entropy.

Properly assessing uncertainty is important 
for foundational quantum
mechanics~\cite{LQ17, KW16a,KW16b} and for quantum information and
communication~\cite{SVA16,SDW17,YBPM17,GT09};
variance is closer than entropy for practical quantum mechanics,
a driving motivation behind research into sum-uncertainty
relations (SURs), which deliver state-independent lower
bounds~\cite{Hol82,PS07,RL08,BLW14,MP14,MP15,BP16,XWZ+17,MBP17}.
Here we discuss 
SURs by showing connections with the
algebras of observables, with examples of the Weyl-Heisenberg
$\mathfrak{wh}$, special unitary~$\mathfrak{su}(n)$ and $\mathfrak{su}(1,1)$
and generally semi-simple compact algebras
\togli{Extending SURs to general ~$\mathfrak{su}(n)$ has implications for nuclear physics~\cite{RW10} 
  and quantum information~\cite{BdGS02}.}  thereby extending the
  range to applications of SURs in areas
  such as~\cite{RW10,IW94} and
  quantum information~\cite{BdGS02} where $\mathfrak{u}(n)$ or
  $\mathfrak{su}(n)$ symmetries are prevalent.

Indeed, single-photon multi-path quantum optical interferometry provides a convenient way to realize $SU(n)$ 
symmetry~\cite{YMK86,RMBB94,dGDS18,CHM+16}
with the experimental signature obtained via sampling photodetection of the photon emerging from each of the $n$ output ports, both by direct detection and 
by adding special post-processing interferometers at the output followed by photodetection. Photodetection sampling statistics obtained in these ways yield 
uncertainties from estimates second-order cumulants for the distributions and, through this process, our uncertainty relations can be empirically tested.  Such 
uncertainty relations are important for assessing the ultimate limits of quantum interferometry~\cite{DJK15} .
Our approach lays a path towards general uncertainty relations with
state-independent lower bounds for arbitrary algebras.

We strongly emphasize that our results refer to the ``preparation
uncertainty'' \cite{KW16a,KW16b} and not to the ``measurement
uncertainty''.  The former refers to the variance of the outcomes of
measurements of different observables performed on different systems
prepared in identical states. The latter, which has been the subject
of a lively debate
recently~\cite{Oza03,Oza04,BKOE13,BLW14,MSP16,RVH17}, refers to
the relation between errors and post-measurement disturbance in an
apparatus.  We underline {these} are very different notions
\cite{Per02,Oza03,BKOE13,BLW14,MSP16,RVH17}, {even if} this difference is
often obscured in the literature.

As our relations are based on the sum of variances, they easily relate
more than two observables and possess a simple physical interpretation
as the diagonal of the uncertainty volume,
depicted in Fig.~\ref{f:fig}.
\begin{figure}
\epsfxsize=1.\hsize\leavevmode\epsffile{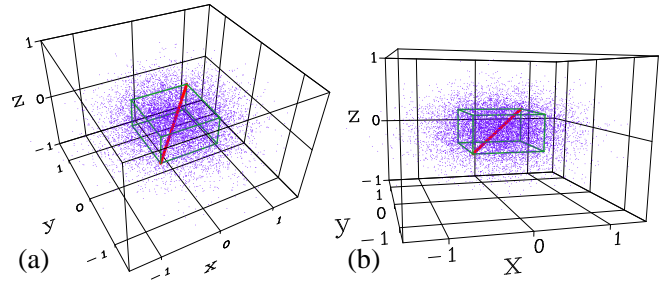}
\caption{%
  Sum of variances is a measure of total uncertainty.
  Given a (green) box with the uncertainties as edges,
  the sum of variances is the squared length of the (red) diagonal.
  [Here 30000 (blue) points with Gaussian distribution corresponding to
  $\Delta x=\Delta y=1/2,\;\Delta z=1/4$ are plotted as an illustration.%
  }
\labell{f:fig}
\end{figure}
The uncertainty relations that we derive from algebraic properties are stated below,
with the explanation and derivation to follow.
Some of our relations are prior knowledge and some are new.
Each algebra is 
defined by its commutator relation,
given by Eq.~(\ref{eq:xp}) for~$\mathfrak{wh}$ and by
\begin{align}
	\left[J_x,J_y\right]=&\text{i}J_z & 
	\left[J_y,J_z\right]=\text{i}J_x& 
	\left[J_z,J_x\right]=\text{i}J_y\nonumber\\
	\left[K_x,K_y\right]=&-\text{i}K_z&
	\left[K_y,K_z\right]=\text{i}K_x&
	\left[K_z,K_x\right]=\text{i}K_y
\end{align}
for $\mathfrak{su}(2)$ and $\mathfrak{su}(1,1)$, respectively.
For semisimple compact Lie algebras \cite {Cor84, Sla81, GIL08}, we use
$\{e_i\}$ in an operator  basis with diagonal Killing form,
$\{\lambda_i\}$ as group irrep labels,
$\ket{\Lambda}$ the {integral dominant} weight and
$\ket{\delta}$ the Weyl root.

We claim the following tight state-independent SURs:
\begin{align}
\label{eq:whsur}
	\mathfrak{wh}&:
		\Delta x^2+\Delta p^2\geq 1,  \mbox{
                \cite{Hei27}}\\
\label{eq:su11sur}
	\mathfrak{su}(1,1)&:
		\Delta K_x^2+\Delta K_y^2-\Delta K_z^2\geq \kappa,
\end{align}
$\kappa $ being Bargmann index and{, more generally, for} the semisimple compact case
\begin{equation}
\label{eq:sccsur}
	\frac{1}{2}\sum_i\Delta e_i^2\geq2\langle\Lambda|\delta\rangle,
\end{equation}
{with specialization to}
\begin{align}
\label{eq:su2sur}
	\mathfrak{su}(2)&:
		\Delta J_x^2+\Delta J_y^2+\Delta J_z^2 
			\geq j, \mbox{ e.g.~in \cite{Del77,DSW15,HT03}}\\
\label{eq:su3sur}
	\mathfrak{su}(3)&:
		\frac{1}{2}\sum_i\Delta e_i^2
			\geq2\left(\lambda_1+\lambda_2\right),\\
\label{eq:su4sur}
	\mathfrak{su}(4)&:
		\frac{1}{2}\sum_i\Delta e_i^2
			\geq3\lambda_1+4\lambda_2+3\lambda_3,\\
\label{eq:su5sur}
	\mathfrak{su}(5)&:
		\frac{1}{2}\sum_i\Delta e_i^2\geq4\lambda_1
		+6\lambda_2+6\lambda_3+4\lambda_4.
\end{align}
{All} these state-independent lower bounds {are} given {functions} that
depend only {linearly} on the choice of irreducible representation (irrep). 
As discussed later, there are states for which the equality holds.
A geometric intuition for relations (\ref{eq:su11sur}-\ref{eq:su5sur}) follows from
Pythagoras' theorem:
the left-hand-side is the squared length of the
diagonal of a ``box'' with uncertainties as edges shown in Fig.~\ref{f:fig}.
These edges are a measure of the ``total'' uncertainty and {bounded} from below by a positive constant.

We begin by developing our  approach {to} the SUR
based on the familiar~$\mathfrak{wh}(1)$ algebra~(\ref{eq:xp})
treated by Heisenberg~\cite{Hei27}.
Heisenberg's uncertainty relation~(\ref{eq:xp}) \cite{Hei27,Rob29} follows from
\begin{equation}
	\left(\Delta x-\Delta p\right)^2\geq0\implies\Delta x^2+\Delta p^2\geq 2\Delta x\Delta p\geq 1,
\end{equation}
which incorporates both the sum~(\ref{eq:whsur}) and the Heisenberg
product $\Delta x\Delta p\geq1/2$~(\ref{eq:xp}) relations in the same
expression.  Our focus is on the sum relation, and we now {rederive}
this SUR by another approach.

We express~$\mathfrak{wh}$ in terms of lowering~$a:=(x+\text{i}p)/\sqrt{2}$
and raising~$a^\dag$ ladder operators so
\begin{equation}
\label{eq:wh}
	\mathfrak{wh}=\operatorname{span}\left\{a,a^\dag,\mathds{1}\right\},\;
	[a,a^\dag]=\mathds{1},
\end{equation}
with the ``weight'', or number, operator denoted $n=a^\dag a$.
The (Fock) eigenstates
$\left\{\ket{m};m\in\mathbb{N}\right\}$
satisfy
\begin{equation}
\label{eq:Fock}
	n\ket{m}
		=m\ket{m},\;
	a^\dag\ket{m}
		=\sqrt{m+1}\ket{m+1}.
\end{equation}
Henceforth, a general arbitrary state is expressed as a sum $\ket{\psi}=\sum_m\psi_m\ket{m}$ over the weights~$\{m\}$ determined
by diagonal operators for the algebra being studied.  If some weights
are repeated, the sum extends over the orthogonal states of the same
weights. For $\mathfrak{wh}$,
\begin{equation}
	\Delta x^2+\Delta p^2
		=2\langle a^\dag a\rangle+1-{\langle x\rangle}^2-{\langle p\rangle}^2.
\labell{sv1}
\end{equation}
is bounded by first considering
\begin{equation}
	\langle x\rangle+\text{i}\langle p\rangle
		=\sqrt{2}\sum_m\nu_m,\;
	\nu_m:=\sqrt{m+1}\psi_{m+1}\psi_m^*.
\labell{xm}
\end{equation}
The Cauchy-Schwarz inequality yields
\begin{align}
	\left\langle x\right\rangle^2+\langle p\rangle^2
		=&2\left|\sum_{n=0}^\infty\nu_n\right|^2
\labell{xp}\\
		\leq&2\sum_{n=0}^\infty(n+1)\left|\psi_{n+1}\right|^2
			\sum_{n'=0}^\infty\left|\psi_{n'}\right|^2
		=2\langle a^\dag a\rangle,\;\nonumber
\end{align}
which proves Eq.~(\ref{eq:whsur}),
and the ``lowest-weight state'' $|0\rangle$ saturates this bound.

Next we apply this~$\mathfrak{wh}$ {technique} 
to the ubiquitous $\mathfrak{su}(2)$ algebra,
pertinent to spin-like systems with $2j\in\mathbb{N}$.
For~$J_\pm=J_x\pm\text{i}J_y$ the $\mathfrak{su}(2)$ raising/lowering operators,
$\mathfrak{su}(2)=\operatorname{span}\left\{J_+,J_-,J_z\right\}$ such that
\begin{equation}
\label{eq:su2algebra}
	\left[J_+,J_-\right]=2J_z,\;
	\left[J_z,J_\pm\right]=\pm J_\pm,
\end{equation}
The eigenstates $\{\ket{m}; 0\leq m\le 2j\}$ of the weight operator $J_z$,  
satisfying
$J_z\ket{m}=(m-j)\ket{m}$
and
\begin{equation}
	J_+\ket{m}
		=\sqrt{(m+1)(2j-m)}\ket{m+1},
\end{equation}
form a basis for the $(2j+1)$-dimensional irrep of~$\mathfrak{su}(2)$ with
\begin{equation}
	C_2
		=J_z^2+\tfrac{1}{2}\left({J_+J_-+J_-J_+}\right)
		=J_z^2+J_x^2+J_y^2
		=c_2\mathds{1}
\end{equation}
and eigenvalue $\langle c_2\rangle =j(j+1)$.
Then
\begin{equation}
\label{eq:su2sum}
	\Delta J_x^2+\Delta J_y^2+\Delta J_z^2
		=
		c_2-\displaystyle\sum_{i=x,y,z}\langle J_i\rangle^2.
\end{equation}
This sum~(\ref{eq:su2sum}) can be bounded, analogous to the
$\mathfrak{wh}$ case, by expanding for an arbitrary state
\begin{equation}
	\ket{\psi}=\sum\psi_m\ket{m}.
\end{equation}

For
\begin{equation}
	\mu_m
		:=\psi^*_{m+1}\psi_{m}\sqrt{(m+1)(2j-m)},
\end{equation}
we have
\begin{equation}
	\left\langle J_z\right\rangle=\sum_{m=0}^{2j}m|\psi_m|^2-j\,,\;
	\left\langle J_x\right\rangle+\text{i}\left\langle J_y\right\rangle=\sum_{m=0}^{2j-1}\mu_m\, ,
\end{equation}
which leads to
\begin{align}
	&\left\langle J_x\right\rangle^2+\left\langle J_y\right\rangle^2
		=\left|\sum_{m=0}^{2j-1}\mu_m\right|^2
				\nonumber\\
		&\leq\left(\sum_{m=0}^{2j-1}|\psi_{m+1}|^2(m+1)\right) 
		\left(\sum_{m'=0}^{2j-1}\left|\psi_{m'}\right|^2(2j-m') \right)
		 		\nonumber\\
		&=2j\sum_m|\psi_m|^2m
			-\Big(\sum_m|\psi_{m}|^2m\Big)^2
\labell{p23}
\end{align}
using the Cauchy-Schwarz inequality.
As
\begin{equation}
	\left\langle J_z\right\rangle^2=j^2-2j\sum_m|\psi_{m}|^2m
		+\Big(\sum_{m}|\psi_m|^2m\Big)^2
\end{equation}
we obtain the desired $\mathfrak{su}(2)$ uncertainty relation~(\ref{eq:su2sur}).
Next we see {how} this approach robustly extends to the noncompact case.

Closely related to $\mathfrak{su}(2)$ is the non-compact
$\mathfrak{su}(1,1)=\operatorname{span}\left\{K_+,K_-,K_z\right\}$
with ladder operators
\begin{equation}
	K_\pm
		=K_x\pm\text{i}K_y
\end{equation}
and commutation relations
\begin{align}
	\left[K_+,K_-\right]=-2K_z\, ,\; 
\left[K_z,K_\pm\right]=\pm K_\pm\, ,
\end{align}
where the operators $K_{x,y,z}$
are self-adjoint on Hilbert space. 
$K_z$ eigenstates~$\{\ket{m}\}$,
such that
\begin{equation}
	K_z\ket{m}
		=(m+\kappa)\ket{m},\;
	m,\kappa\geq0,
\end{equation}
form a basis for the infinite-dimensional unitary irrep $\kappa$.
We restrict our discussion to irreps of the positive discrete series, where the representation label common in physics are 
\begin{equation}
	\kappa=1/2, 1, 3/2, \ldots.
\end{equation}
The analysis {also} applies to the two limits of discrete series with labels $\kappa=1/4,3/4$.
The eigenvalue~$m$ is discrete; 
continuous $m$~\cite{DDM01} is a topic for future investigation.
The $\mathfrak{su}(1,1)$ raising operators satisfies
\begin{equation}
\label{eq:K+-}
	K_+\ket{m}=\sqrt{(m+1)(2\kappa+m)}\ket{m+1},
\end{equation}
and $K_-=K_+^\dag$.
Evidently the ladder of $\ket{m}$ states is unbounded above,
but  the $K_z$ eigenstate~$\ket{m=0}$
with eigenvalue $\kappa$
is annihilated by $K_-$.
The quadratic Casimir operator is
\begin{equation*}
	C_2
		=K_z^2-\tfrac{1}{2}\left({K_+K_-+K_-K_+}\right)
		=K_z^2-K_x^2-K_y^2
		=c_2\mathds{1}\,
\end{equation*}
with $c_2=\kappa(\kappa-1)$.
The sum of variances is
\begin{align}
	\Delta K_x^2+\Delta K_y^2&+\Delta K_z^2
		\geq\Delta K_x^2+\Delta K_y^2-\Delta K_z^2\nonumber \\
		&= \langle K_x\rangle^2 + \langle K_y\rangle^2 -  \langle K_z\rangle^2-c_2. \label{su11boundedsum}
\end{align}
Assuming all sums converge, we bound this 
by expanding an arbitrary state $\ket{\psi}=\sum\psi_m\ket{m}$ to obtain
\begin{equation}
	\left\langle K_z\right\rangle
		=\sum_{m=0}^\infty m\left|\psi_m\right|^2+\kappa,\;
	\left\langle K_x\right\rangle+\text{i}\left\langle K_y\right\rangle
		=\sum_{m=0}^\infty\lambda_m
\end{equation}
with
$\lambda_m:=\psi^*_{m+1}\psi_m\sqrt{(m+1)(2\kappa+m)}$,
whence
\begin{align}
	\left\langle K_x\right\rangle^2&+\left\langle K_y\right\rangle^2
		=\left|\sum_{m=0}^\infty\lambda_m\right|^2\nonumber\\
		&\leq\left(\sum_{m=0}^\infty\left|\psi_{m+1}\right|^2(m+1)\right)
			\left(\sum_{m'=0}^\infty\left|\psi_{m'}\right|^2(2\kappa+m')\right)
						\nonumber\\
		&=\left(\sum_{m=0}^\infty\left|\psi_{m}\right|^2m\right)
			\left(\sum_{m'=0}^\infty\left|\psi_{m'}\right|^2(2\kappa+m')\right)
					\nonumber\\
		&=2\kappa\sum_m|\psi_m|^2m
			+\left(\sum_m|\psi_m|^2m\right)^2, \label{eq:avgKxplusKysquared}
\end{align}
using the Cauchy-Schwarz inequality.
As
\begin{equation}
	\left\langle K_z\right\rangle^2
		=\kappa^2+2\kappa\sum_m|\psi_m|^2m
			+\left(\sum_m|\psi_m|^2m\right)^2,  \label{eq:avgKzsquared}
\end{equation}
we can recover the right side of Eq.(\ref{su11boundedsum}) 
by subtracting Eq.(\ref{eq:avgKxplusKysquared}) from Eq.(\ref{eq:avgKzsquared}) to 
obtain
\begin{equation}
	\left\langle K_z\right\rangle^2-\left\langle K_x\right\rangle^2-\left\langle K_y\right\rangle^2\geq\kappa^2.
\end{equation}
and thus the desired SUR~(\ref{eq:su11sur}). 
Actually we have proven the inequality (\ref{su11boundedsum}),
which is even stronger than desired.

We have so far discussed three cases of SURs,
all based on ladder operator relations and Casimir operators,
although the $\mathfrak{wh}$ case has a trivial Casimir operator, namely $\mathds{1}$.
We now have the tools to investigate more general cases involving semisimple Lie algebras.

Consider a compact semi-simple rank-$r$ Lie algebra
\begin{equation}
\label{eq:gspan}
	\mathfrak{g}
		=\operatorname{span}
			\left\{
				e_k, e_m;
					k\in\left\{1,\ldots,r\right\},
						m\in\left\{r+1,\dots,\ell\right\}
							\right\}
\end{equation}
with~$e_k$ a diagonal Cartan element and~$e_m$ a nondiagonal operator.
For $\mathfrak{su}(2)$ this would be the Hermitian basis with $e_1=J_z$, and $e_{2,3} = J_{x,y}$.
The $\ell-r$ operators are combinations of raising
and lowering operators so, crucially, have null expectation value on
any eigenstate of the Cartan elements, i.e., on any state of definite weight.

The Casimir operator~$C_2$ and its state-independent eigenvalue~$c_2$ are 
\begin{equation}
\label{eq:c2}
	C_2=\frac{1}{2}\sum_{k=1}^\ell e_k^2,\;
	c_2=2\langle\Lambda\ket{\delta}+\langle\Lambda\ket{\Lambda},\;
	\ket{\Lambda}:=\sum_{i=1}^r\lambda_i\ket{w_i}
\end{equation}
with~$\ket{\Lambda}$ the 
highest-weight for the irrep $\Lambda=(\lambda_1,\dots,\lambda_r)$,
$\ket{w_i}$  the $i^\text{th}$ fundamental weight {and}~$\ket{\delta}$ the Weyl root.
The Weyl root is half the sum of all positive roots as 
detailed in \cite{Cor84} or 
\cite{Sla81}.
Scalar products~(\ref{eq:c2}) are computed
with a metric matrix~$G$ \cite{Sla81}:
for
\begin{equation}
	\ket{\mu}
		:=\sum_i\mu_i\ket{w_i},\;
	\langle\mu|\tau\rangle
		=\mu\cdot G\cdot \tau.
\end{equation}

The sum of the variances of all $\{ e_i\}$ is
\begin{equation}
\label{eq:sumo}
	\frac{1}{2}\sum_{k=1}^\ell\Delta e_k^2
		=\langle  C_2\rangle-\frac{1}{2}\sum_{k=1}^\ell\langle  e_k\rangle^2
		=c_2-\frac{1}{2}\sum_{k=1}^r\langle e_k\rangle^2,
\end{equation}
where,
in the last equality,
we assume the system state $\ket{\lambda}$ is an eigenstate of the~$r$ Cartan operators
so that
\begin{equation}
	\langle  e_m\rangle
		=\bra{\lambda} e_m\ket{\lambda}=0
\end{equation}
for
$m>r$ due to the action of the raising and lowering operators.
For the weight~$\ket{\lambda}$,
\begin{equation}
\label{eq:esum}
	\frac{1}{2}\sum_{k=1}^r\langle e_k\rangle^2
		=\langle\lambda\ket{\lambda}
		\leq\langle\Lambda\ket{\Lambda}
\end{equation}
where the upper bound is attained for the highest-weight state.
Combining Eqs.~(\ref{eq:sumo}) and~(\ref{eq:esum}) yields
\begin{equation}
\label{eq:sscur}
	\frac{1}{2}\sum_{k=1}^\ell\Delta e_k^2\geq c_2
			-\langle\Lambda\ket{\Lambda}
		=2\langle\Lambda\ket{\delta}, 
\end{equation}
which is the desired SUR~(\ref{eq:sccsur}) for semisimple compact Lie
algebras.  Moreover, the uncertainty-sum relation is tight as the
inequality is saturated by the highest-weight state~$\ket{\Lambda}$,
its Weyl-reflected images and any state in the group orbit of
$\ket{\Lambda}$, i.e.any coherent state~\cite{Per86}.

We now demonstrate the value of Eq.~(\ref{eq:sscur}) through its application to examples of compact unitary algebras,
namely $\mathfrak{su}(3)$, $\mathfrak{su}(4)$ and $\mathfrak{su}(5)$.
For~$\mathfrak{su}(3)$,
Eq.~(\ref{eq:su3sur}) follows immediately from Eq.~(\ref{eq:sscur}) using
\begin{equation}
	G_{SU(3)}
		=\frac{1}{3} 
		\begin{pmatrix}
			 2&1\\1&2 \\
		\end{pmatrix}.
\label{su3Gmatrix}
\end{equation}
The Hermitian basis for the defining,
i.e., 3-dimensional $(\lambda_1,\lambda_2)=(1,0)$,
irrep of~$\mathfrak{su}(3)$ is
\begin{align}
	A_-
	=&\begin{pmatrix}
	0&-\text{i} &0 \\
	+\text{i}&0&0 \\
 0&0&0
\end{pmatrix},
A_+
=\begin{pmatrix}
 0&1&0 \\
 1&0&0 \\
 0&0&0
\end{pmatrix},
B_-
=\begin{pmatrix}
 0&0&0 \\
 0&0&-\text{i} \\
 0&\text{i} &0
\end{pmatrix},
			\nonumber\\
B_+
=&\begin{pmatrix}
 0&0&0 \\
 0&0&1 \\
 0&1&0
\end{pmatrix},
C_-=
\begin{pmatrix}
 0&0&-\text{i} \\
 0&0&0 \\
 \text{i}&0&0
\end{pmatrix},\;
C_+
	=\begin{pmatrix}
		0&0&1\\0&0&0\\1&0&0
\end{pmatrix},
			\nonumber\\
	h_1=&\operatorname{diag}(1,-1,0),\;
	h_2=\operatorname{diag}(1,1,-2).
\label{10irrep}
\end{align}
The Killing form is $2\times \unit$ whereas the quadratic Casimir operator
\begin{equation*}
	 C_2
		=\frac{1}{2}\left(
			A_+^2+ A_-^2+B_+^2+ B_-^2+ C_+^2+ C_-^2
				+h_1^2+h_2^2\right)
\end{equation*}
{has} eigenvalue 
\begin{equation} 
	c_2\left(\lambda_1,\lambda_2\right)
		=\frac{2}{3}\left(\lambda_1^2+\lambda_2^2
			+3\left[\lambda_1+\lambda_2\right]
				+\lambda_1\lambda_2\right)
\end{equation} 
for irrep $(\lambda_1,\lambda_2)$.
For the $(1,0)$ irrep~(\ref{10irrep}),
$C_2=\frac{8}{3}\unit$.

\togli{Now we verify inequality~(\ref{eq:su3sur}) for a different $\mathfrak{su}(3)$ irrep,
namely the (8-dimensional) adjoint irrep~$(1,1)$.
In this case}

With this, we easily verify the lower uncertainty bound
\begin{align}
	\frac{1}{2}
		&\big[\Delta(A_+)^2+\Delta(A_-)^2+\Delta(B_+)^2+\Delta(B_-)^2  \nonumber\\
			&+\Delta(C_+)^2+\Delta(C_-)^2
				+\Delta(h_1)^2+\Delta(h_2)^2\big]
					\nonumber \\
	:&=\frac{1}{2}\sum_i (\Delta \tilde e_i)^2 
	\geq2(\lambda_1+\lambda_2),
\label{su3bound}
\end{align}
\togli{which confirms that the general formula~(\ref{eq:sscur})
gives the correct inequality~(\ref{eq:su3sur}).}

We generalize this procedure to $\mathfrak{su}(4)$
and~$\mathfrak{su}(5)$, yielding (\ref{eq:su4sur}) and (\ref{eq:su5sur}) respectively,
and confirm our procedure
for irreps $(\lambda_1,\lambda_2,\lambda_3)$
and $(\lambda_1,\lambda_2,\lambda_3,\lambda_4)$, respectively.
For $\mathfrak{su}(4)$,
we obtain the Gell-Mann matrices $\Lambda_{1-15}$ in Appendix A following Stover's procedure~\cite{Sto}
to obtain
\begin{align}
	c_2&(\lambda_1,\lambda_2,\lambda_3)
		=\frac{1}{4}\big(3\lambda_1^2
			+2\left[2 \lambda_2+\lambda_3
				+6\right]\lambda_1\nonumber\\&
			+4 \lambda_2^2+4 \lambda_2\left[\lambda_3+4\right]
				+3\lambda_3\left[\lambda_3+4\right]\big).
\end{align}
The lower bound~(\ref{eq:su4sur}) is successfully obtained with each~$e_i$ replaced by~$\Lambda_i$
so our expression 
is confirmed for this $\mathfrak{su}(4)$ irrep.

For $\mathfrak{su}(5)$ we have the $5\times 5$ Gell-Mann
matrices $\Lambda'_{1-24}$ given in Appendix A and we obtain
\begin{align} 
	c_2&(\lambda_1,\lambda_2,\lambda_3,\lambda_4)= {\tfrac{2}{5}}
	[2 \lambda_1^2+3 \lambda_2 \lambda_1
		+2 \lambda_3 \lambda_1+\lambda_4 \lambda_1+
\nonumber \\&
10 \lambda_1
+3 \lambda_2^2+3 \lambda
   _3^2+2 \lambda_4^2+15 \lambda_2+4 \lambda_2 \lambda_3+15
   \lambda_3+2 \lambda_2
\nonumber \\&
\lambda_4+3 \lambda_3 \lambda_4+10
   \lambda_4].
\end{align} 
Replacing each~$e_i$ in inequality~(\ref{eq:su5sur}) by the Gell-Mann matrix~$\Lambda_i$
given in the Appendix A confirms that the
SUR~(\ref{eq:su5sur}) holds for this~$\mathfrak{su}(5)$ irrep.  {We}
note that the bound is the same for conjugate irreps, {\it v.g.} the
$\mathfrak{su}(3)$ irreps $(\lambda_1,\lambda_2)$ and
$(\lambda_2,\lambda_1)$ have the same bound, with similar symmetry for
conjugate representations holding for $\mathfrak{su}(4)$ and
$\mathfrak{su}(5)$ irreps.

Whereas Eq.~(\ref{eq:sscur}) is an equality on the sums of all
  variances, one can also obtain various inequalities involving sums
  over a restricted set of variances. There are also some special cases of our inequalities that appear in
  \cite{,GMT+06, VHE+11}. T\'oth {\it et al} \cite{GKG+07} have given inequalities involving
  variances and expectation values of $\mathfrak{su}(2)$ for detecting
  bound entanglement in spin systems. One can reproduce their proofs for
  $\mathfrak{su}(n)$ (Appendix B) and find that a
  violation of
 \begin{align}
(N-1)\sum_{k=1}^{n-1} (\Delta e_k)^2\ge \sum_{m}\langle e_m^2 \rangle -2(n-1)N \label{inqent}\, ,
\end{align}
implies entanglement for $N$ particles, each in the fundamental representation $(1,\ldots,0)$ of $\mathfrak{su}(n)$.  In (\ref{generaltoth2c}), the sum over $k$ is a sum of the variances
of the $r=n-1$ elements in the Cartan subalgebra of $\mathfrak{su}(n)$, whereas the sum over $m$ is over the remaining elements not in the Cartan subalgebra. 

As a simple example of application of Eq.~(\ref{generaltoth2c}), fix $n$ and consider the $n$-fold coupling of the fundamental of 
$\mathfrak{su}(n)$, i.e. the $n$-fold coupling of $(1,0,\ldots,0)$.  The scalar irrep $(0,0,...\ldots,0)$ occurs once in this decomposition.  
 The states of the scalar irrep in this $n$-fold coupling are determinants in the $n$ states.  
 In $\mathfrak{su}(2)$, this would be the coupling of two spin-1/2 particles to the entangled $s=0$ singlet state. 
 For $\mathfrak{su}(3)$, with basis states $\vert 100\rangle, \vert 010\rangle,\vert 001\rangle$, the scalar that appears as the 3-particle coupling is the (entangled) determinant state
$$
\vert\psi\rangle = {\cal N} \left\vert \begin{array}{ccc}
\vert 100\rangle_1& \vert 010\rangle_1&\vert 001\rangle_1\\
\vert 100\rangle_2& \vert 010\rangle_2&\vert 001\rangle_2\\
\vert 100\rangle_3& \vert 010\rangle_3&\vert 001\rangle_3
\end{array}\right\vert\, ,
$$
where ${\cal N}$ is a normalization.
Clearly since this state is in $(0,0)$ of $\mathfrak{su}(3)$, $\Delta e_k=0$ and $\langle e_m^2\rangle=0$ for all $k$ and $m$ for this state.  Thus, our inequality ~(\ref{generaltoth2c}) becomes
$$
0 \ge 0 -2\times 2\times 3 = -12
$$
and so is clearly violated, correctly implying that $\vert\psi\rangle$ is entangled.

Finally, T\'oth {\it et al}
also obtain an inequality containing the sums of all the variances: 
this is nothing but our Eq.~(\ref{eq:sscur}) for states in the irrep $(N,0,\ldots)$,
which are not the highest weight state,
its reflection, or a coherent state for this irrep. This generalizes
the entanglement detection results of \cite{GKG+07} to $\mathfrak{su}(n)$.

In conclusion, we have presented a class of state-independent tight
SURs based on algebraic properties, and our scheme shows how to
generalize to other algebras.  Inequalities~(\ref{eq:whsur})
and~(\ref{eq:su2sur}) were known
previously~\cite{DSW15,HT03,Del77,RL08,SB16,ACF+16,FS16} as is
(implicitly) inequality~(\ref{eq:su11sur})~\cite{Per86}, but bounds
were not explicitly stated nor was their common algebraic origin from
a similar derivation.  Furthermore the state-independent nature of the
tight lower bound was not investigated.  Instead previous analyses
focused on their connection with algebraic coherent
states~\cite{Per86}.  Different relations for state-independent
variance-based uncertainty relations were known explicitly only for
qubits~\cite{BLW14,AAHB16,Oza04}. Whereas state-independent
  uncertainty relations were traditionally connected to entropic
  uncertainty relations, our results show how one can obtain them also
  for variance-based ones.

The result of Eq.~(\ref{eq:sscur}) exploits the relation between the SUR and the quadratic Casimir operator when the Killing form is diagonal,
which is an easily generalizable notion including
to infinite-dimensional irreps,
but the state saturating this lower bound might not be normalizable.
Some work needs to be done, such as dealing with continuous irreps, verifying SURs for various irreps
and generalizing to other algebras.
Some aspects of our work were known before but not in a unified, explicit, purely algebraic approach as done here.

\acknowledgments HdG and BCS each acknowledge NSERC support.  LM
acknowledges funding from Unipv ``Blue sky'' project-grant BSR1718573
and the FQXi foundation grant FQXi-RFP-1513.  NS acknowledges support
from the University of Calgary Eyes High postdoctoral fellowship
program.  

\section{APPENDIX A: Gell-Mann matrices for $\mathfrak{su}(4)$ and $\mathfrak{su}(5)$}

The $4\times 4$ Gell-Mann matrices are
\begin{align*} 
\Lambda_1
=&\begin{pmatrix}
0&1&0&0 \\
 1&0&0&0 \\
 0&0&0&0 \\
 0&0&0&0
\end{pmatrix},\;
 \Lambda_2
=\begin{pmatrix}
 0&0&1&0 \\
 0&0&0&0 \\
 1&0&0&0 \\
 0&0&0&0
\end{pmatrix},\\
\Lambda_3
=&\begin{pmatrix}
 0&0&0&1 \\
 0&0&0&0 \\
 0&0&0&0 \\
 1&0&0&0
\end{pmatrix},\;
\Lambda_4
=\begin{pmatrix}
 0&0&0&0 \\
 0&0&1&0 \\
 0&1&0&0 \\
 0&0&0&0
\end{pmatrix},\\
\Lambda_5
=&\begin{pmatrix}
 0&0&0&0 \\
 0&0&0&1 \\
 0&0&0&0 \\
 0&1&0&0
\end{pmatrix},\;
\Lambda_6
=\begin{pmatrix}
 0&0&0&0 \\
 0&0&0&0 \\
 0&0&0&1 \\
 0&0&1&0
\end{pmatrix},\\
\Lambda_7
=&\begin{pmatrix}
 0&-\text{i} &0&0 \\
 \text{i}&0&0&0 \\
 0&0&0&0 \\
 0&0&0&0
\end{pmatrix},\;
\Lambda_8
=\begin{pmatrix}
 0&0&-\text{i} &0 \\
 0&0&0&0 \\
 \text{i}&0&0&0 \\
 0&0&0&0
\end{pmatrix},
\end{align*}
and
\begin{align*}
\Lambda_9
=&\begin{pmatrix}
 0&0&0&-\text{i} \\
 0&0&0&0 \\
 0&0&0&0 \\
 \text{i}&0&0&0
\end{pmatrix},\;
\Lambda_{10}
=\begin{pmatrix}
 0&0&0&0 \\
 0&0&-\text{i} &0 \\
 0&\text{i} &0&0 \\
 0&0&0&0
\end{pmatrix},\\
\Lambda_{11}
=&\begin{pmatrix}
 0&0&0&0 \\
 0&0&0&-\text{i} \\
 0&0&0&0 \\
 0&\text{i} &0&0
\end{pmatrix},\;
\Lambda_{12}
=\begin{pmatrix}
 0&0&0&0 \\
 0&0&0&0 \\
 0&0&0&-\text{i} \\
 0&0&\text{i} &0
\end{pmatrix},\\
\Lambda_{13}
=&\begin{pmatrix}
 1&0&0&0 \\
 0&-1&0&0 \\
 0&0&0&0 \\
 0&0&0&0
\end{pmatrix},\;
\Lambda_{14}
=\begin{pmatrix}
 \frac{1}{\sqrt3} &0&0&0 \\
 0&\frac{1}{\sqrt3} &0&0 \\
 0&0&-\frac{2}{\sqrt3} &0 \\
 0&0&0&0
\end{pmatrix},\\
\Lambda_{15}
=&\begin{pmatrix}
 \frac{1}{\sqrt{6}} &0&0&0 \\
 0&\frac{1}{\sqrt{6}} &0&0 \\
 0&0&\frac{1}{\sqrt{6}} &0 \\
 0&0&0&-\sqrt{\frac{3}{2}}
\end{pmatrix}
\end{align*}
The $5\times 5$ Gell-Mann matrices are.
 \begin{align*}
\Lambda'_1
=&\begin{pmatrix}
0&1&0&0&0 \\
 1&0&0&0&0 \\
 0&0&0&0&0 \\
 0&0&0&0&0 \\
 0&0&0&0&0
\end{pmatrix},\;
\Lambda'_2
=\begin{pmatrix}
 0&0&1&0&0 \\
 0&0&0&0&0 \\
 1&0&0&0&0 \\
 0&0&0&0&0 \\
 0&0&0&0&0
\end{pmatrix},\\
\Lambda'_3
=&\begin{pmatrix}
 0&0&0&1&0 \\
 0&0&0&0&0 \\
 0&0&0&0&0 \\
 1&0&0&0&0 \\
 0&0&0&0&0
\end{pmatrix},\;
\Lambda'_4
=\begin{pmatrix}
 0&0&0&0&1 \\
 0&0&0&0&0 \\
 0&0&0&0&0 \\
 0&0&0&0&0 \\
 1&0&0&0&0
\end{pmatrix},\\
\Lambda'_5
=&\begin{pmatrix}
 0&0&0&0&0 \\
 0&0&1&0&0 \\
 0&1&0&0&0 \\
 0&0&0&0&0 \\
 0&0&0&0&0
\end{pmatrix},\;
\Lambda'_6
=\begin{pmatrix}
 0&0&0&0&0 \\
 0&0&0&1&0 \\
 0&0&0&0&0 \\
 0&1&0&0&0 \\
 0&0&0&0&0
\end{pmatrix},\\
\Lambda'_7
=&\begin{pmatrix}
 0&0&0&0&0 \\
 0&0&0&0&1 \\
 0&0&0&0&0 \\
 0&0&0&0&0 \\
 0&1&0&0&0
\end{pmatrix},\;
\Lambda'_8
=\begin{pmatrix}
 0&0&0&0&0 \\
 0&0&0&0&0 \\
 0&0&0&1&0 \\
 0&0&1&0&0 \\
 0&0&0&0&0
\end{pmatrix},
\end{align*}
and
\begin{align*}
\Lambda'_9
=&\begin{pmatrix}
 0&0&0&0&0 \\
 0&0&0&0&0 \\
 0&0&0&0&1 \\
 0&0&0&0&0 \\
 0&0&1&0&0
\end{pmatrix},\;
\Lambda'_{10}
=\begin{pmatrix}
 0&0&0&0&0 \\
 0&0&0&0&0 \\
 0&0&0&0&0 \\
 0&0&0&0&1 \\
 0&0&0&1&0
\end{pmatrix},\\
\Lambda'_{11}
=&\begin{pmatrix}
 0&-\text{i} &0&0&0 \\
 \text{i}&0&0&0&0 \\
 0&0&0&0&0 \\
 0&0&0&0&0 \\
 0&0&0&0&0
\end{pmatrix},\;
\Lambda'_{12}
=\begin{pmatrix}
 0&0&-\text{i} &0&0 \\
 0&0&0&0&0 \\
 \text{i}&0&0&0&0 \\
 0&0&0&0&0 \\
 0&0&0&0&0
\end{pmatrix},\\
\Lambda'_{13}
=&\begin{pmatrix}
 0&0&0&-\text{i} &0 \\
 0&0&0&0&0 \\
 0&0&0&0&0 \\
 \text{i}&0&0&0&0 \\
 0&0&0&0&0
\end{pmatrix},\;
\Lambda'_{14}
=\begin{pmatrix}
 0&0&0&0&-\text{i} \\
 0&0&0&0&0 \\
 0&0&0&0&0 \\
 0&0&0&0&0 \\
 \text{i}&0&0&0&0
\end{pmatrix},\\
\Lambda'_{15}
=&\begin{pmatrix}
 0&0&0&0&0 \\
 0&0&-\text{i} &0&0 \\
 0&\text{i} &0&0&0 \\
 0&0&0&0&0 \\
 0&0&0&0&0
\end{pmatrix},\;
\Lambda'_{16}
=\begin{pmatrix}
 0&0&0&0&0 \\
 0&0&0&-\text{i} &0 \\
 0&0&0&0&0 \\
 0&\text{i} &0&0&0 \\
 0&0&0&0&0 \\
\end{pmatrix},
\end{align*}
and
\begin{align*}
\Lambda'_{17}
=&\begin{pmatrix}
 0&0&0&0&0 \\
 0&0&0&0&-\text{i} \\
 0&0&0&0&0 \\
 0&0&0&0&0 \\
 0&\text{i} &0&0&0
\end{pmatrix},\;
\Lambda'_{18}
=\begin{pmatrix}
 0&0&0&0&0 \\
 0&0&0&0&0 \\
 0&0&0&-\text{i} &0 \\
 0&0&\text{i} &0&0 \\
 0&0&0&0&0
\end{pmatrix},\\
\Lambda'_{19}
=&\begin{pmatrix}
 0&0&0&0&0 \\
 0&0&0&0&0 \\
 0&0&0&0&-\text{i} \\
 0&0&0&0&0 \\
 0&0&\text{i} &0&0
\end{pmatrix},\;
\Lambda'_{20}
=\begin{pmatrix}
 0&0&0&0&0 \\
 0&0&0&0&0 \\
 0&0&0&0&0 \\
 0&0&0&0&-\text{i} \\
 0&0&0&\text{i} &0
\end{pmatrix},\\
\Lambda'_{21}
=&\begin{pmatrix}
 1&0&0&0&0 \\
 0&-1&0&0&0 \\
 0&0&0&0&0 \\
 0&0&0&0&0 \\
 0&0&0&0&0
\end{pmatrix},\;
\Lambda'_{22}
=\begin{pmatrix}
 \frac{1}{\sqrt3} &0&0&0&0 \\
 0&\frac{1}{\sqrt3} &0&0&0 \\
 0&0&-\frac{2}{\sqrt3} &0&0 \\
 0&0&0&0&0 \\
 0&0&0&0&0
\end{pmatrix},\\
\Lambda'_{23}
=&\begin{pmatrix}
 \frac{1}{\sqrt{6}} &0&0&0&0 \\
 0&\frac{1}{\sqrt{6}} &0&0&0 \\
 0&0&\frac{1}{\sqrt{6}} &0&0 \\
 0&0&0&-\sqrt{\frac{3}{2}} &0 \\
 0&0&0&0&0
\end{pmatrix},\\
\Lambda'_{24}
=&\begin{pmatrix}
 \frac{1}{\sqrt{10}} &0&0&0&0 \\
 0&\frac{1}{\sqrt{10}} &0&0&0 \\
 0&0&\frac{1}{\sqrt{10}} &0&0 \\
 0&0&0&\frac{1}{\sqrt{10}} &0 \\
 0&0&0&0&-2 \sqrt{\frac{2}{5}}
\end{pmatrix}
\end{align*}

\section{APPENDIX B: Derivation of the inequality (\ref{inqent}) in the main text} 
All separable states of $N$ systems, with properties described by
operators in an $\mathfrak{su}(n)$ algebra, satisfy the following
inequality. Namely a violation of the inequality implies that the
systems are entangled:
\begin{align}
(N-1)\sum_{k=1}^{r} (\Delta e_k)^2\ge \sum_{m=r+1}^{\ell} \langle e_m^2 \rangle -2(n-1)N \label{generaltoth2c}\, ,
\end{align}
where the sum over $k$ is a sum over the $r=n-1$ commuting elements in the Cartan subalgebra of $\mathfrak{su}(n)$,
and the sum of $m$ is a sum over the remaining non-diagonal operators in $\mathfrak{su}(n)$.  The operators in
$\mathfrak{su}(n)$ are normalized so that $\hbox{Tr}(e_a^\dagger e_b)=2\delta_{ab}$, as per Eq.(36) for $\mathfrak{su}(3)$.  We derive Eq.(\ref{generaltoth2c}) explicitly for $\mathfrak{su}(3)$ and discuss the specific parts of the
derivation that will generalize to $\mathfrak{su}(n)$.

Denote 
$\langle e_\alpha^i\rangle=\lambda_\alpha^i$, where $\alpha$ labels a Gell-Mann matrix and $i$ the particle.  The collective operators $e_\alpha =\sum_{i=1}^N e_{\alpha}^i$.  The Cartan elements are $e_1$ and $e_2$.  Then:
\begin{align}
(N-1)\left((\Delta e_1)^2+(\Delta e_2)^2\right)-\sum_{m=3}^8 \langle e_m^2 \rangle +4N\ge 0
\label{generalrelation2c}
\end{align}
Using
\begin{align}
(\Delta e_1)^2&=\langle e_1^2\rangle - \langle \lambda_1\rangle^2\, ,\\
&= \langle (\sum_i e_1^i) (\sum_j e_1^j)\rangle -
\langle (\sum_i e_1^i)\rangle\langle (\sum_j e_1^j)\rangle\, .
\end{align}
If the states are factorizable ({\it i.e.} separable) then the average values satisfy:
\begin{align}
(\Delta e_1)^2&=\sum_i \langle(e_1^i)^2\rangle  -\sum_i (\lambda_1^i)^2\, .
\end{align}
Doing the same for $e_2$ and summing gives
\begin{align}
&\left((\Delta e_1)^2+(\Delta e_2)^2\right)\\
&= \sum_i \left[\langle(e_1^i)^2\rangle+\langle(e_2^i)^2\rangle\right] -
\sum_i \left[(\lambda_1^i)^2+(\lambda_2^i)^2\right]  \label{Cartansum}.
\end{align}
From the explicit expression of the Gell-Mann matrices, one finds
\begin{align}
(e_1^i)^2+(e_2^i)^2=\frac{4}{3}1_{3\times 3},
\end{align}
so that 
\begin{align}
&(N-1)\left((\Delta e_1 )^2+(\Delta e_2)^2\right)\\
&= \frac{4}{3}N(N-1) -
(N-1)\left[\sum_i (\lambda_7^i)^2+(\lambda_8^i)^2\right] \label{lhs}.
\end{align}
Next, 
\begin{align}
\sum_{m}\langle e_m^2\rangle -4N &=\sum_\beta 
\langle  (\sum_i e_m^i)(\sum_j e_m^j)\rangle \, ,\\
&= \sum_{m} \left[\sum_i \langle \left( e_m^i\right)^2 \rangle +\sum_{i\ne j}  
\langle e_m^i e_m^j\rangle \right] -4N .
\end{align}
One easily verifies that, for fixed $i$
\begin{align}
\sum_{m} \left( e_m^i\right)^2 =  4 \times \unit_{3\times 3} \label{eq:Cartansquared},
\end{align}
so that we now have
\begin{align}
&\sum_{m}\langle e_m^2\rangle -4N
= \sum_m \sum_{i\ne j} 
\lambda_m^i\lambda_m^j\\
&\qquad \le \sum_m \left(\sum_i\lambda_m^i\right)^2 -\sum_m \sum_i\left(\lambda_m^i \right)^2\, ,\\
&\qquad \le \sum_m N\sum_i \left(\lambda_m^i\right)^2 -\sum_m \sum_i\left(\lambda_m^i \right)^2\, ,\\
&\qquad \le (N-1)\sum_m \sum_i \left(\lambda_m^i\right)^2 \, .\label{rhs}
\end{align}
Subtracting Eqs.(\ref{rhs}) from (\ref{lhs}) yields
\begin{align}
&(N-1)\left((\Delta e_1)^2+(\Delta e_2)^2\right)-\sum_{m}\langle e_m^2\rangle -4N \nonumber \\
&\quad \ge \frac{4}{3}N(N-1) -
(N-1)\sum_{\alpha=1}^8 \sum_i \left(\lambda_\alpha^i\right)^2.
\end{align}
Finally, one readily verifies that, for fixed $i$, 
\begin{align}
\displaystyle\sum_{\alpha=1}^8 (\lambda_\alpha^i)^2=\frac{4}{3} \label{sunBlochsphere},
\end{align} 
so that 
\begin{align}
(N-1)\sum_{\alpha=1}^8 \sum_i \left(\lambda_\alpha^i\right)^2 = \frac{4}{3}N(N-1),
\end{align}
from which Eq.(\ref{generalrelation2c}) follows.

For $\mathfrak{su}(4)$, Eq.(\ref{Cartansum}) becomes
\begin{align}
(e_1^i)^2+(e_2^i)^2+(e_3^i)^2=\frac{3}{2}\unit_{4\times 4},
\end{align}
while Eq.(\ref{sunBlochsphere}) gives $\displaystyle\sum_{\alpha=1}^{15} (\lambda_\alpha^i)^2=\frac{3}{2}$.  Moreover, the factor $4N$ is replaced by $6N$ so that Eq.(\ref{generaltoth2c}) follows.

For $\mathfrak{su}(5)$, Eq.(\ref{Cartansum}) becomes
\begin{align}
(e_1^i)^2+(e_2^i)^2+(e_3^i)^2+(e_4^i)^2=\frac{8}{5}\unit_{5\times 5},
\end{align}
while Eq.(\ref{sunBlochsphere}) gives $\displaystyle\sum_{\alpha=1}^{24} (\lambda_\alpha^i)^2=\frac{8}{5}$.  This time,
the factor $4N$ is replaced by $8N$ and Eq.(\ref{generaltoth2c}) follows. This argument still holds for mixed states since, by convexity of variance, $ ({\Delta e_1})^2_{(\rho)} > \sum_i p_i ({\Delta e_1})^2_{(\ket{\psi_i})}$ for $\rho=\ket\psi_i\bra\psi$.

Finally, we can conjugate the entire algebra by any (global) $n\times n$ unitary matrix $U$ since, by Eq.(\ref{eq:Cartansquared}) the sum $\sum_m (e_m^i)^2$ of squares of Cartan elements is proportional to the unit matrix and thus invariant under transformation by $U$.   
\bibliography{unc-resub}
\end{document}